\documentclass[showpacs,showkeys,amsmath,amssymb,aps,prb,preprint]{revtex4-1}
\usepackage[justification=RaggedRight]{caption}
\usepackage{subcaption}
\usepackage{graphicx}
\usepackage{dcolumn}
\usepackage{bm}
\usepackage{epstopdf}
\usepackage{color}

\begin{document}

\preprint{APS/123-QED}
\title{Interpenetrating Graphene Networks: Three-dimensional Node-line Semimetals with Massive Negative Linear Compressibilities}
\author{Yangzheng Lin$^1$}
\author{Zhisheng Zhao$^{1,2}$}
\author{Timothy A. Strobel$^1$} \email{tstrobel@carnegiescience.edu}
\author{R. E. Cohen$^{1,3}$}
\affiliation{$^1$Extreme Materials Initiative, Geophysical Laboratory, Carnegie Institution of Washington, 5251 Broad Branch Road NW, Washington, DC 20015, USA}
\affiliation{$^2$State Key Laboratory of Metastable Materials Science and Technology, Yanshan University, Qinhuangdao 066004, China}
\affiliation{$^3$Department of Earth- and Environmental Sciences, Ludwig Maximilians Universit\"at, Munich 80333, Germany}

\date{\today}

\begin{abstract}

We investigated the stability and mechanical and electronic properties of fifteen metastable mixed $sp^2$-$sp^3$ carbon allotropes in the family of interpenetrating graphene networks (IGNs) using density functional theory (DFT) within the generalized gradient approximation (GGA). IGN allotropes exhibit non-monotonic bulk and linear compressibilities before their structures irreversibly transform into new configurations under large hydrostatic compression. The maximum bulk compressibilities vary widely between structures and range from 3.6 to 306 TPa$^{-1}$. We find all the IGN allotropes have negative linear compressibilities with maximum values varying from -0.74 to -133 TPa$^{-1}$. The maximal negative linear compressibility of Z33 (-133 TPa$^{-1}$ at 3.4 GPa) exceeds previously reported values at pressures higher than 1.0 GPa. IGN allotropes can be classified as either armchair- or zigzag-type, and these two types of IGNs exhibit different electronic properties. Zigzag-type IGNs are node-line semimetals, while armchair-type IGNs are either semiconductors or node-loop or node-line semimetals. Experimental synthesis of these IGN allotropes might be realized since their formation enthalpies relative to graphite are only 0.1 - 0.5 eV/atom (that of C$_{60}$ fullerene is about 0.4 eV/atom), and energetically feasible binary compound pathways are possible.

\end{abstract}

\maketitle


\section{\label{sec:level1a}Introduction}

Known carbon allotropes with mixed $sp^2$ and $sp^3$ hybridizations are usually amorphous \cite{Suk12}. Multiple carbon crystals with mixed $sp^2$ and $sp^3$ hybridization have been proposed over the past decades \cite{Kuc06,Zhao11,Hu13,Wen13,Dong15,zhao16,Hu161}, although none of them have been convincingly confirmed by experiments. Recent high-resolution transmission electron microscopy (TEM) images, however, suggest that interpenetrating graphene-like networks might exist locally within compressed glassy carbons \cite{Hu162}. Interpenetrating graphene networks (IGNs) are a family of pure carbon allotropes consisting of cross-linked graphene sheets in three dimensions (3D). 3D connectivity of sheets is achieved with $sp^3$ nodes that link graphene sheets and create open pores in the structures. The open pores are rectangular prisms with parallel $sp^3$ carbon chains along the edges, which join $sp^2$ carbon ribbons of variable widths on the sides.

Similar to carbon nanotubes\cite{Milton10}, IGNs can be classified into armchair (A) and zigzag (Z) types according to the $sp^3$ chain and $sp^2$ sheet connectivity along the pore direction (Fig. \ref{fig:Fig01}). There are two pairs of parallel $sp^2$ carbon ribbons on the four sides of IGN pores. In Z-type IGNs, $sp^3$ carbon atoms form six-atom rings with $sp^2$ carbon atoms on both pairs of parallel sides. In A-type IGNs, $sp^3$ carbon atoms form six-atom rings with $sp^2$ carbon atoms on one pair of parallel sides, but form four- and eight-atom rings with $sp^2$ carbon atoms on the others. The carbon ribbons on all sides can be described as a number of armchair or zigzag chains. In this work we designate A-type IGNs as A$ij$ (Fig. \ref{fig:Fig02}a), where $i$ denotes the number of armchair chains in ribbons with four- and eight-atom rings, and $j$ denotes the number of zigzag chains in ribbons with all six-atom rings. Similarly, Z-type IGNs are denoted as Z$ij$ (Fig. \ref{fig:Fig02}b), where $i$ and $j$ denote the number of zigzag chains in the parallel pore ribbons. Z$ij$ and Z$ji$ are identical according to crystallographic symmetry.

Zhao\cite{Zhao12} explored five IGN allotropes (Z11, Z12, Z13, Z23, and Z14) using density functional theory (DFT) and demonstrated that they are energetically metastable with respect to graphite, but exhibit mechanical stability. Later, Jiang \textit{et al.}\cite{Jiang13} studied the mechanical and electronic properties of six kinds of IGN allotropes (A11, A22, A33, Z11, Z22 and Z33) and concluded that these structures are semiconducting and that only Z-type IGNs have negative linear compressibilities. Recently, Chen \textit{et al.}\cite{Chen15} demonstrated that Z11 is not a semiconductor, but actually a semimetal based on detailed numerical computations within DFT and theoretical analysis. In addition to their special mechanical and electronic properties, IGN topologies are calculated to be low-energy metastable structures in high-pressure carbides with composition MC$_6$ (M=metal). For example, it might be possible to obtain Z11 by removing Li or Ca from metastable LiC$_6$\cite{Lin15} or CaC$_6$\cite{Li13}, in a similar fashion to metal removal from zeolite-type silicon structures\cite{Gryko00,Kim15}. In this work, we have discovered six additional energetically competitive and mechanically stable IGN allotropes (A12, A21, A13, A31, A23 and A32) and have investigated the detailed electronic and mechanical properties of the entire IGN family (including 15 structures up to A33 and Z33).

\section{\label{sec:level1b}Computational Methods}

The electronic band structures and the fixed-pressure properties were calculated using density functional theory with the projector augmented-wave (PAW) method\cite{Blochl94,Kresse99} within the Perdew-Burke-Ernzerhof (PBE) generalized gradient approximation (GGA)\cite{Perdew96,Perdew97}. The phonon vibrational frequencies were computed using density-functional perturbation theory (DFPT). All the DFT and DFPT computations were performed using PWSCF and phonon codes as implemented in QUANTUM ESPRESSO\cite{Giannozzi09,Pwscf14}. The plane-wave kinetic-energy cutoff was 80 Ry (1088 eV). In the fixed-pressure relaxations, dense Monkhorst-Pack (MP) k-point meshes were adopted for convergence of the relative enthalpies within several meV per carbon atom.

It is known that different kinds of exchange-correlation functionals in DFT give different lattice parameters and zero-pressure stabilities relative to graphite and diamond \cite{Madsen10,Yu14,Lechner16}. The local-density approximation (LDA) and GGA are the two most widely used approximations for carbon allotropes \cite{Amsler12,Niu12} and many other crystalline systems \cite{Jaffe00,Curtarolo05,Wu06,Zhang10}. We chose GGA as the primary method in this work because it gives better pressure-dependent phase stability predictions than LDA in some crystalline systems \cite{Amorim06,Lin14} and it correctly predicts that graphite is more stable than diamond at ambient pressure \cite{Lechner16}. Although GGA significantly over-predicts the zero-pressure volume of graphite (30 percent larger than LDA), it gives much smaller deviations for the zero-pressure volumes and lattice parameters for IGN allotropes (only 4 percent larger than LDA). Unless otherwise specified, all of the results and discussion in this paper are based on GGA-PBE calculations. For comparison, we also list results from LDA computations in the supporting information (Table SI and Figs. S1-S3).

For each IGN allotrope, we computed the enthalpy and volume ($V$) after relaxation at approximately 50 pressures. We then calculated the bulk and linear compressibilities using the definitions $\beta{_\text{B}}=-(1/V)(\partial V/\partial p)_T$ and $\beta{_\text{L}}=-(1/l)(\partial l/\partial p)_T$ ($V$ is volume and $l$ is lattice distance), respectively, at different pressures. For a given IGN allotrope, we computed the detailed electronic properties and phonon dispersion at one volume corresponding to 0 GPa (1 atm). The k-point meshes used in the electronic properties calculations were very dense compared with those used in structure relaxation so that energy band contacts and Fermi surfaces could be examined in detail.

\section{\label{sec:level1c}Results and Discussion}

\subsection{\label{sec:level2a}Structure and Stability}

\begin{figure*}[!ht]
\centering
\begin{subfigure}[b]{0.9\textwidth}
   \includegraphics[width=0.9\linewidth]{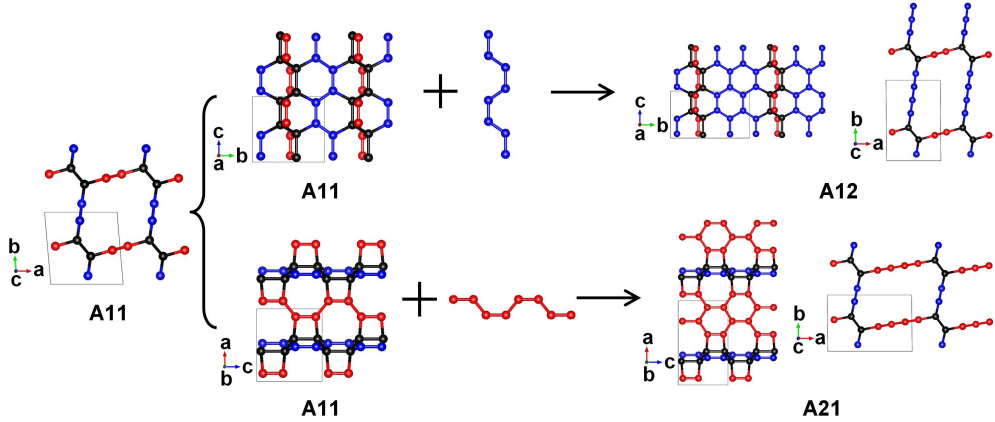}
   \caption{From A11 to A12 and A21}
   \label{fig:Fig01a} 
\end{subfigure}
\begin{subfigure}[b]{0.9\textwidth}
   \includegraphics[width=0.9\linewidth]{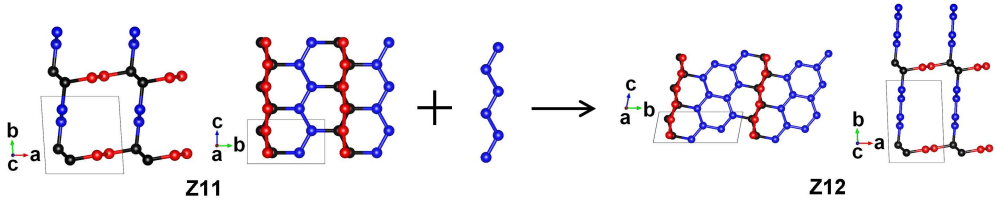}
   \caption{From Z11 to Z12}
   \label{fig:Fig01b}
\end{subfigure}
\caption{Illustration of design principle used to construct IGNs with larger pores. The building units are armchair chains and zigzag chains in A-type and Z-type allotropes, respectively. Non-standard primitive cells are used so that c axes are along the pore and chain directions. The black spheres indicate $sp^3$ hybridized carbon atoms. The red and blue spheres indicate $sp^2$ hybridized carbon atoms along a and b axes, respectively. The same representations are used in next figures, except that we only use blue spheres to represent all $sp^2$ hybridized carbon atoms.}
\label{fig:Fig01} 
\end{figure*}

\begin{figure}[!ht]
\centering
   \begin{subfigure}[b]{0.8\textwidth}
   \includegraphics[width=0.8\linewidth]{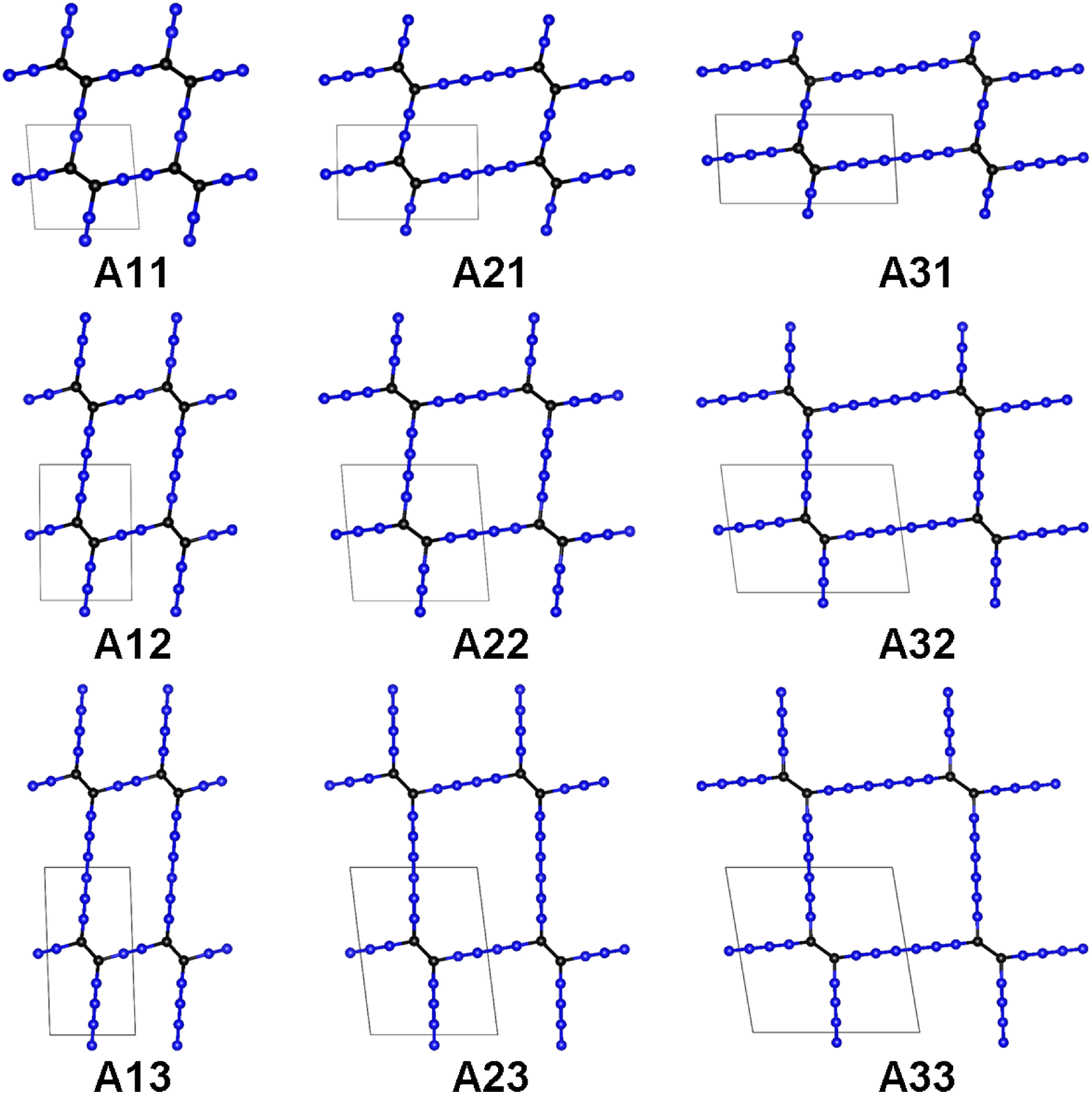}
   \caption{A-type IGNs}
   \label{fig:Fig02a} 
\end{subfigure}
\begin{subfigure}[b]{0.8\textwidth}
   \includegraphics[width=0.8\linewidth]{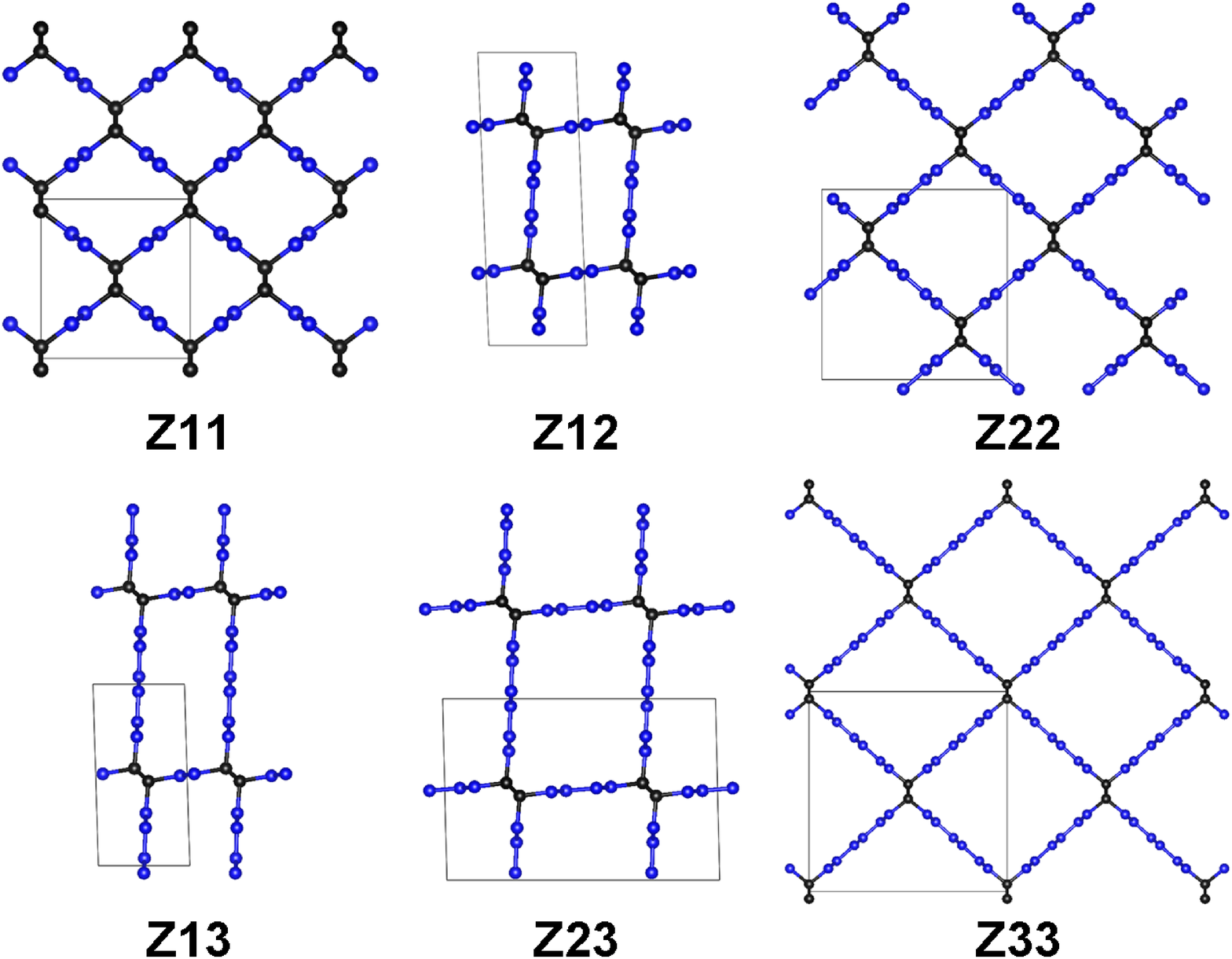}
   \caption{Z-type IGNs}
   \label{fig:Fig02b}
\end{subfigure}
\caption{The crystal structures of (a) A-type and (b) Z-type IGN allotropes.}
\label{fig:Fig02}
\end{figure}

All of the IGN allotropes studied previously by Jiang \textit{et al.}\cite{Jiang13} were symmetric with respect to the pore edge lengths that are normal to the pore direction (i.e., A11, Z11, etc.). Here we expand the number of structures with asymmetric pore lengths originally proposed by Zhao\cite{Zhao12} by adding armchair or zigzag chains to the known structures. For example, we obtained A12 or A21 by inserting one armchair chain to the primitive cell of A11, and obtained Z12 by inserting one zigzag chain to the primitive cell of Z11 (Fig. \ref{fig:Fig01}). In this way, we obtained six additional A-type allotropes (A12, A13, A21, A23, A31, and A32) and three additional Z-type allotropes (Z12, Z13, and Z23) (see in Fig. \ref{fig:Fig02}). We note that an infinite number of structures could be built within this family by increasing the graphene nanoribbon widths. The 15 allotropes examined here have valuable information as to the general trends of properties within the entire IGN family. The detailed structure information of all these 15 IGN allotropes can be found in Table SII in the supporting information. At 0 GPa, the phonon vibrational frequencies in all of the new structures are positive (see Figs. S4-S6 in the supporting information), which indicates that they are all mechanically stable. 

\begin{figure}[!ht]
\centering
\includegraphics[width=0.8\linewidth]{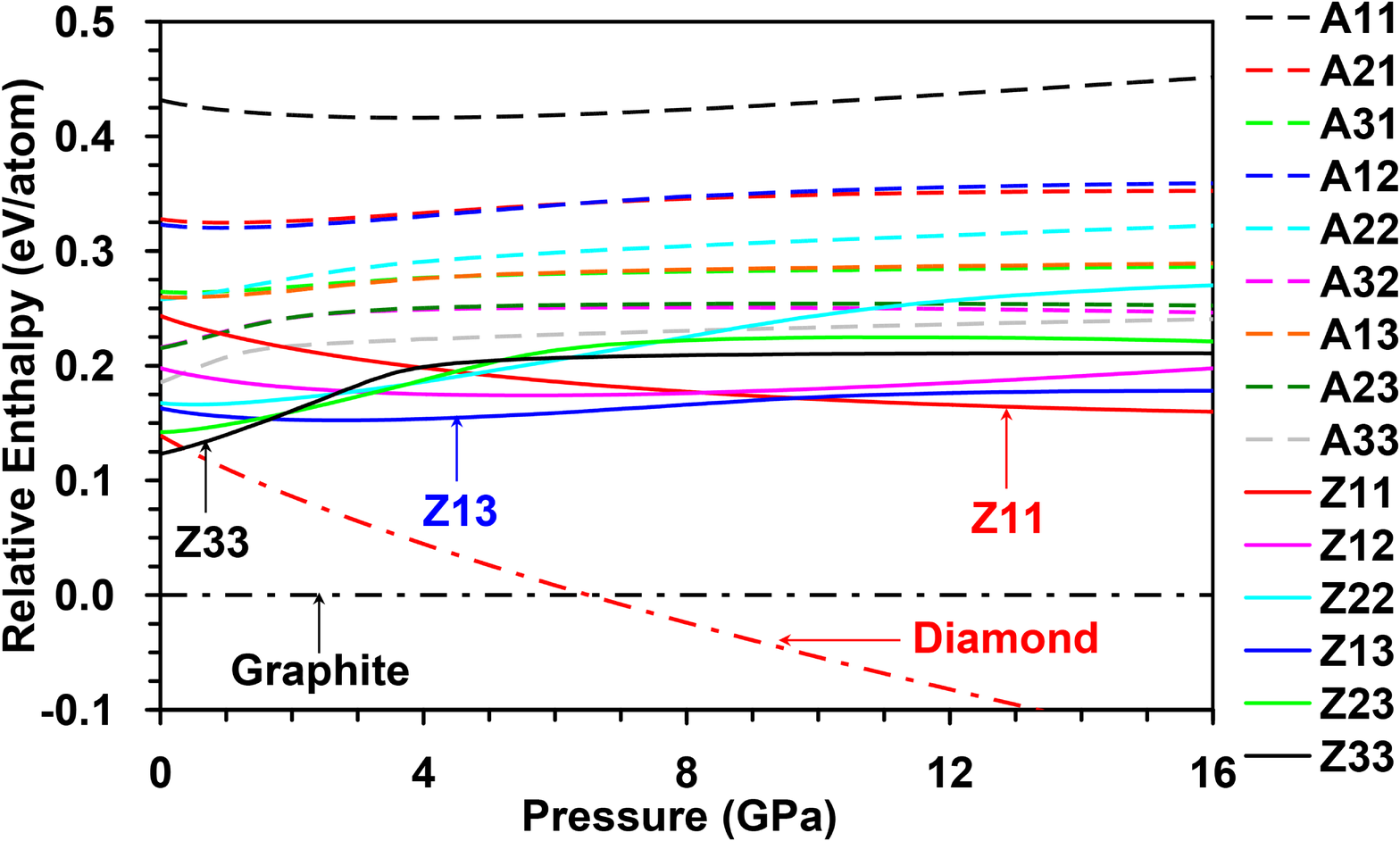}
\caption{Pressure-dependent enthalpies of IGN carbon allotropes and diamond relative to graphite.}
\label{fig:Fig03} 
\end{figure}

From 0 to 16 GPa, the enthalpies all the 15 IGNs relative to graphite are in the range of 0.1 - 0.5 eV/atom (Fig. \ref{fig:Fig03}). At 0 GPa, the formation enthalpy of Z33 is only 0.123 eV/atom, which is smaller than the formation enthalpy of diamond (0.139 eV/atom) at the same pressure. With the same pore length on each side, the Z-type IGNs are energetically more favorable than the A-type. Among all 15 IGNs, Z33, Z13 and Z11 are the most energetically favorable ones at pressures of $<$1.7 GPa, 1.7$-$9.7 GPa and $>$9.7 GPa, respectively. These low formation enthalpies are in a plausible range for experimental synthesis: C$_{60}$, an experimentally known carbon allotrope is metastable with respect to graphite by ~0.4 eV/atom\cite{Cioslowski00}.

The IGN structures are mechanically stable over a broad pressure range during cold compression. We do find, however, that all IGN allotropes transform irreversibly into new structures after compression to very high pressures. At these high pressures, new bonds form between atoms on the neighboring or opposite sides of IGN pores (Figs. S7-S11 in the supporting information), and the formation of these new bonds is irreversible during cold decompression. The maximum pressures for mechanical stability are structure dependent and vary widely amongst the IGN allotropes (see $p_\text{ir}$ in Table \ref{tab:EfCom}). Among all 15 IGNs, A32 loses stability most easily from 18 to 20 GPa, whereas the IGN structures of Z11, Z22 and Z33 remain mechanically stable to pressures higher than 150 GPa. Most of the irreversibly transformed structures (Table SIII in the supporting information) are completely $sp^3$ bonded carbon allotropes except A12, A21, A32 and A23, which still contain a fraction of $sp^2$ bonds. Upon cold-compression, A21 transforms to the \textit{mC}16 structure mentioned by Hu et al. \cite{Hu14}, Z11 transforms to the 3D-(4,0) structure by Zhao et al. \cite{Zhao11}, Z12 transforms to so-called ``M-Carbon''
 \cite{Oganov06,Li09,Zhang11,Botti13}, and Z13 transforms to the \textit{P}2$_1$/\textit{m} structure mentioned by Zhang et al. \cite{Zhang13} The other transformed structures from A11, A12, A13, A22, A23, A31, A32, A33, Z22, Z23 and Z33 are different from any previously reported carbon allotropes \cite{Amsler13,Georgakilas15}, including those listed in the SACADA database \cite{Hoffmann16}.

\subsection{\label{sec:level2b}Bulk and Linear Compressibilities}

\begin{figure}[!ht]
\centering
   \begin{subfigure}[b]{0.8\textwidth}
   \includegraphics[width=0.8\linewidth]{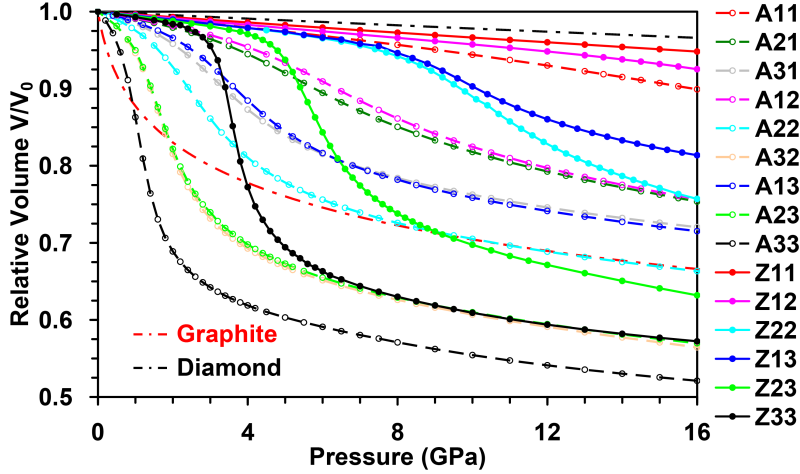}
   \caption{Relative volume}
   \label{fig:Fig04a} 
\end{subfigure}
\begin{subfigure}[b]{0.8\textwidth}
   \includegraphics[width=0.8\linewidth]{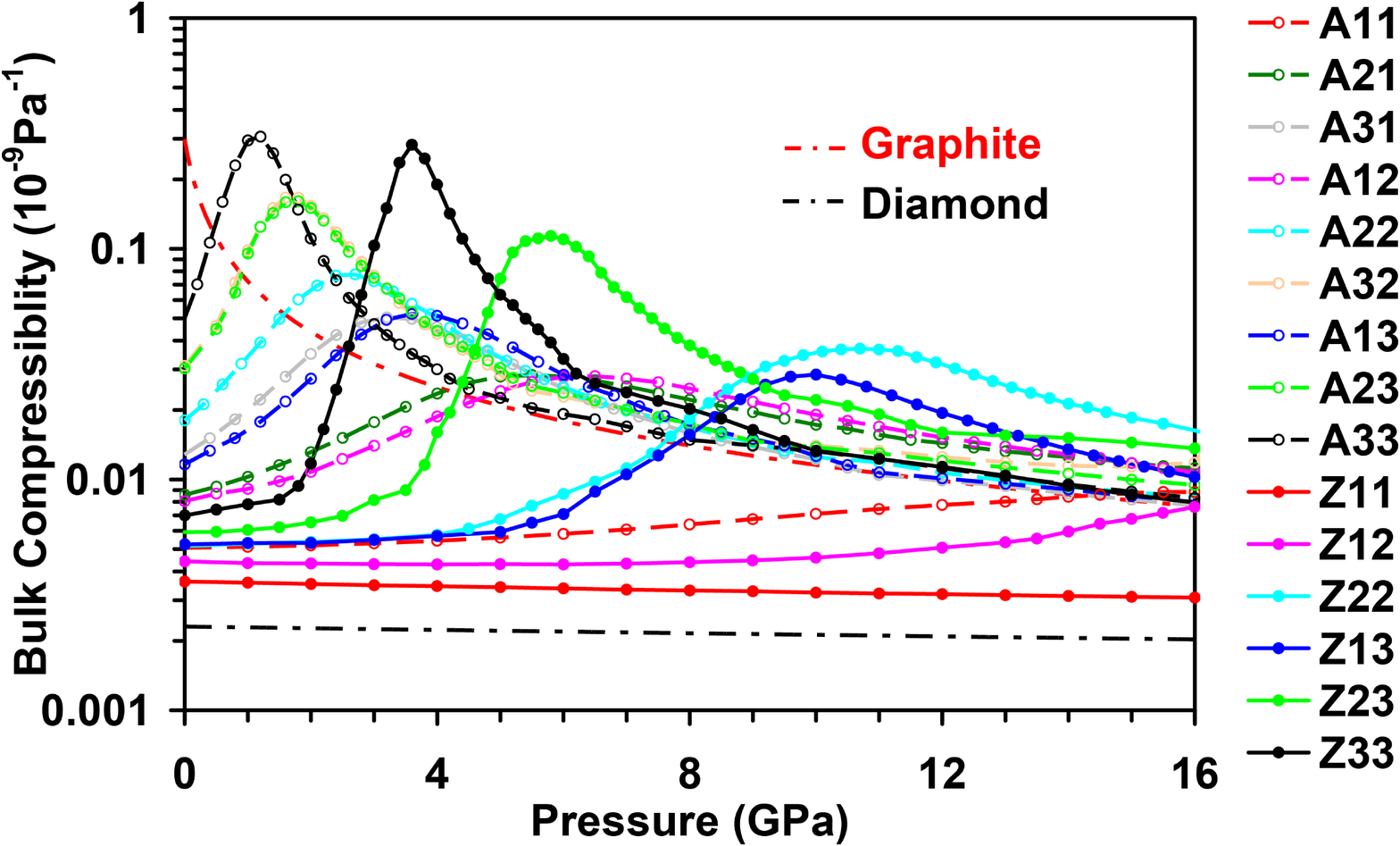}
   \caption{Bulk compressibility}
   \label{fig:Fig04b}
\end{subfigure}
\caption{Pressure-dependent (a) relative volumes and (b) bulk compressibilities of carbon allotropes. }
\label{fig:Fig04}
\end{figure}

\begin{figure}[!ht]
\centering
\begin{subfigure}[b]{0.9\textwidth}
   \includegraphics[width=0.35\linewidth]{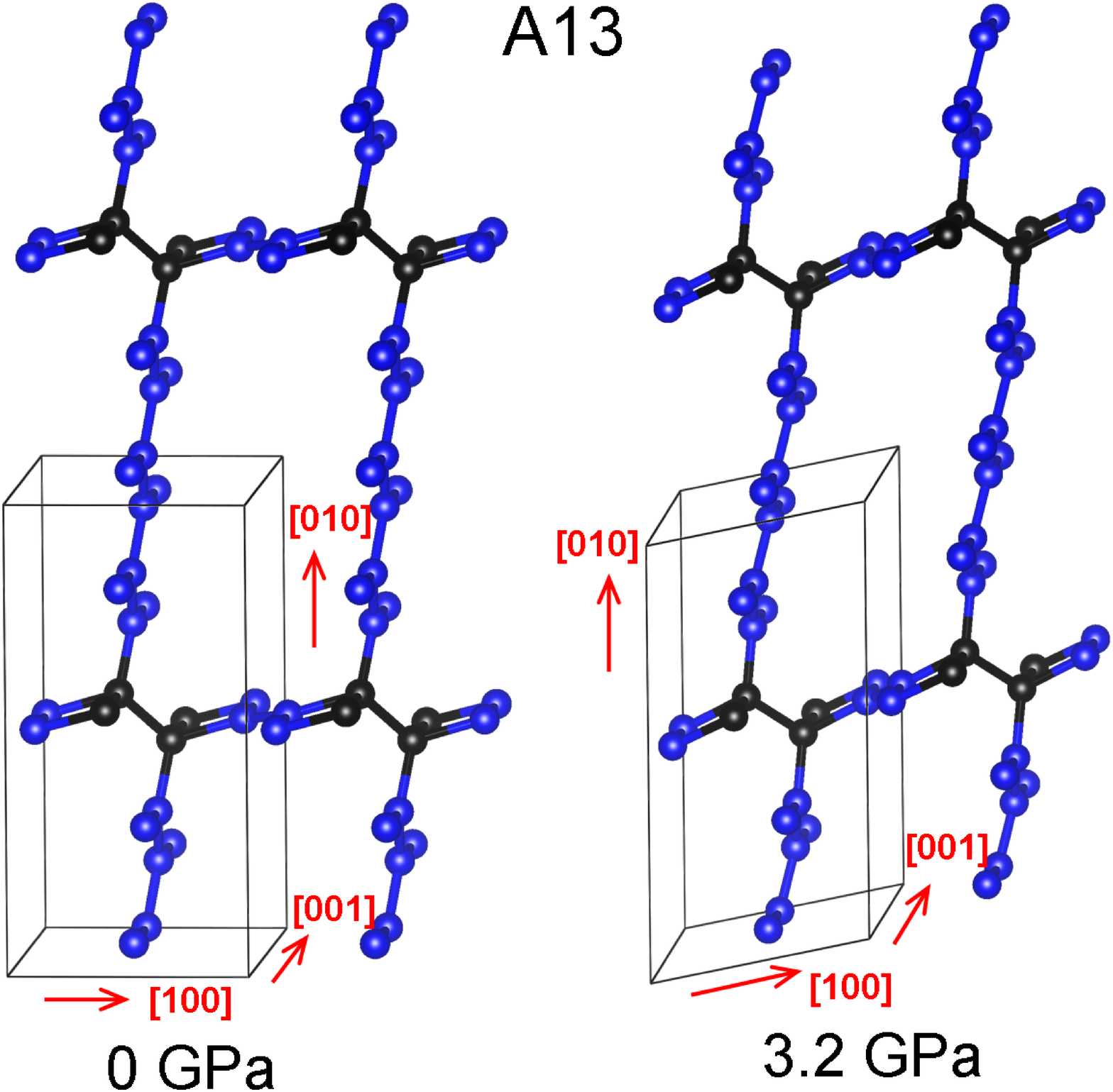}
   \includegraphics[width=0.55\linewidth]{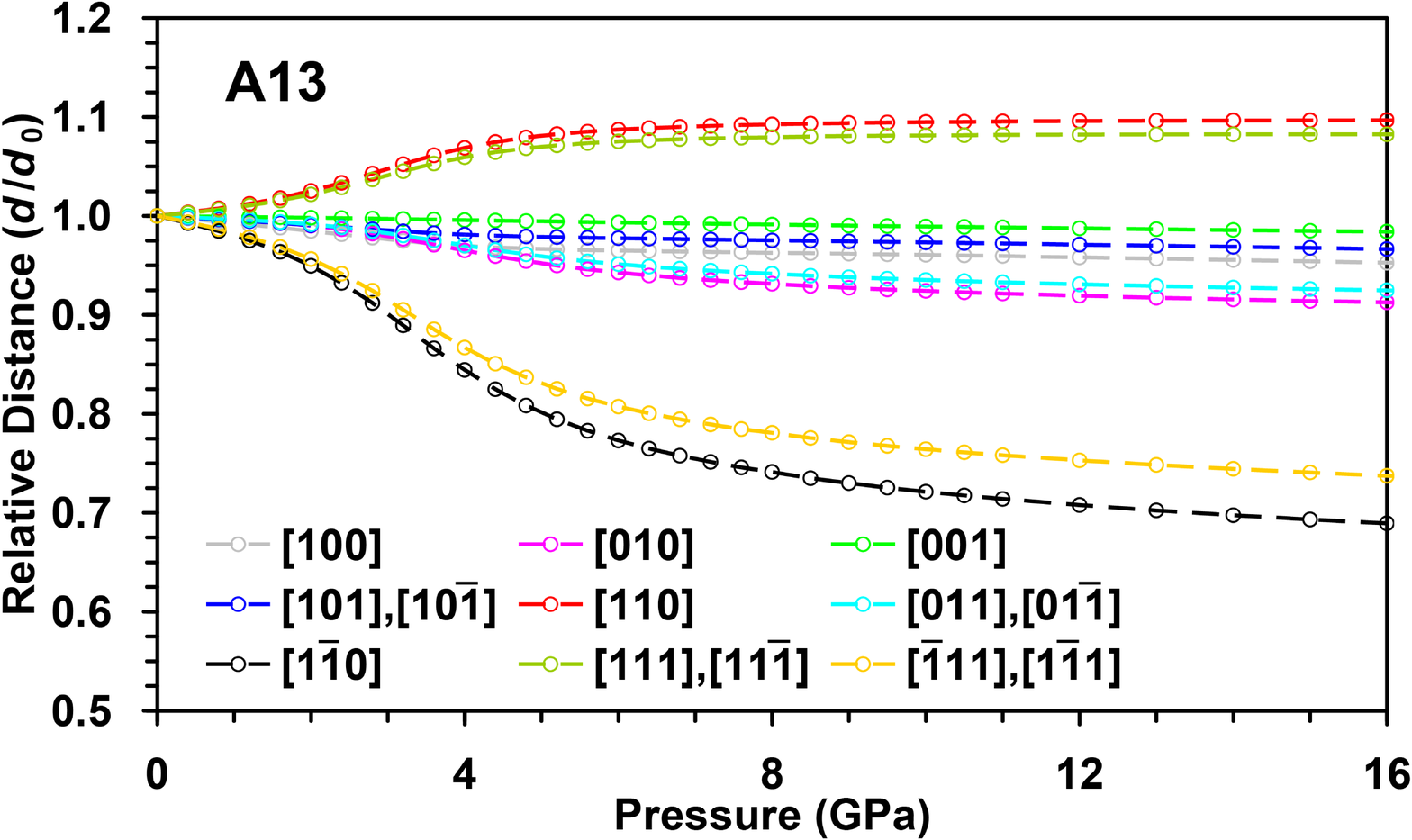}
   \caption{A13}
   \label{fig:Fig05ab} 
\end{subfigure}
\begin{subfigure}[b]{0.9\textwidth}
   \includegraphics[width=0.35\linewidth]{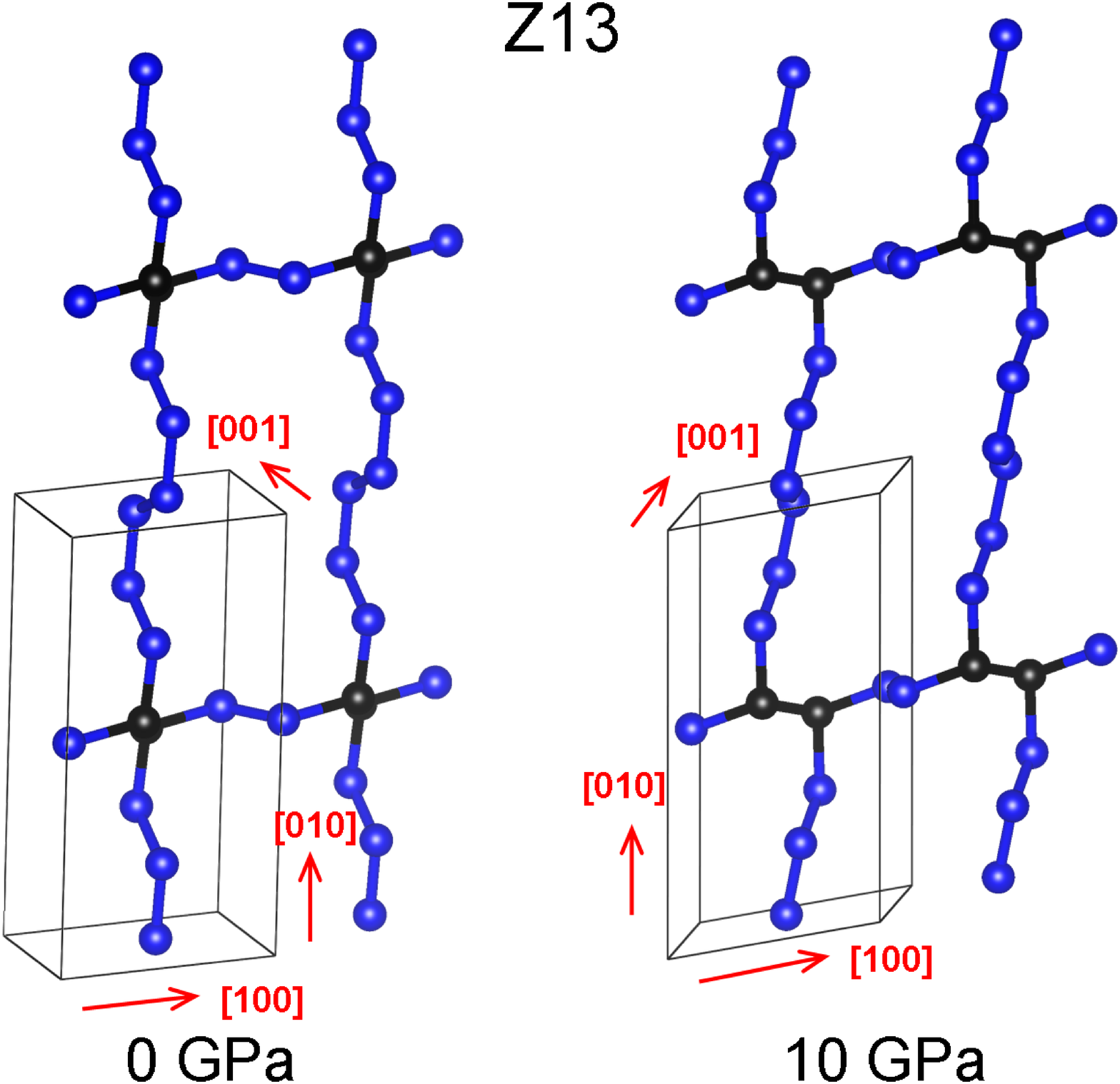}
   \includegraphics[width=0.55\linewidth]{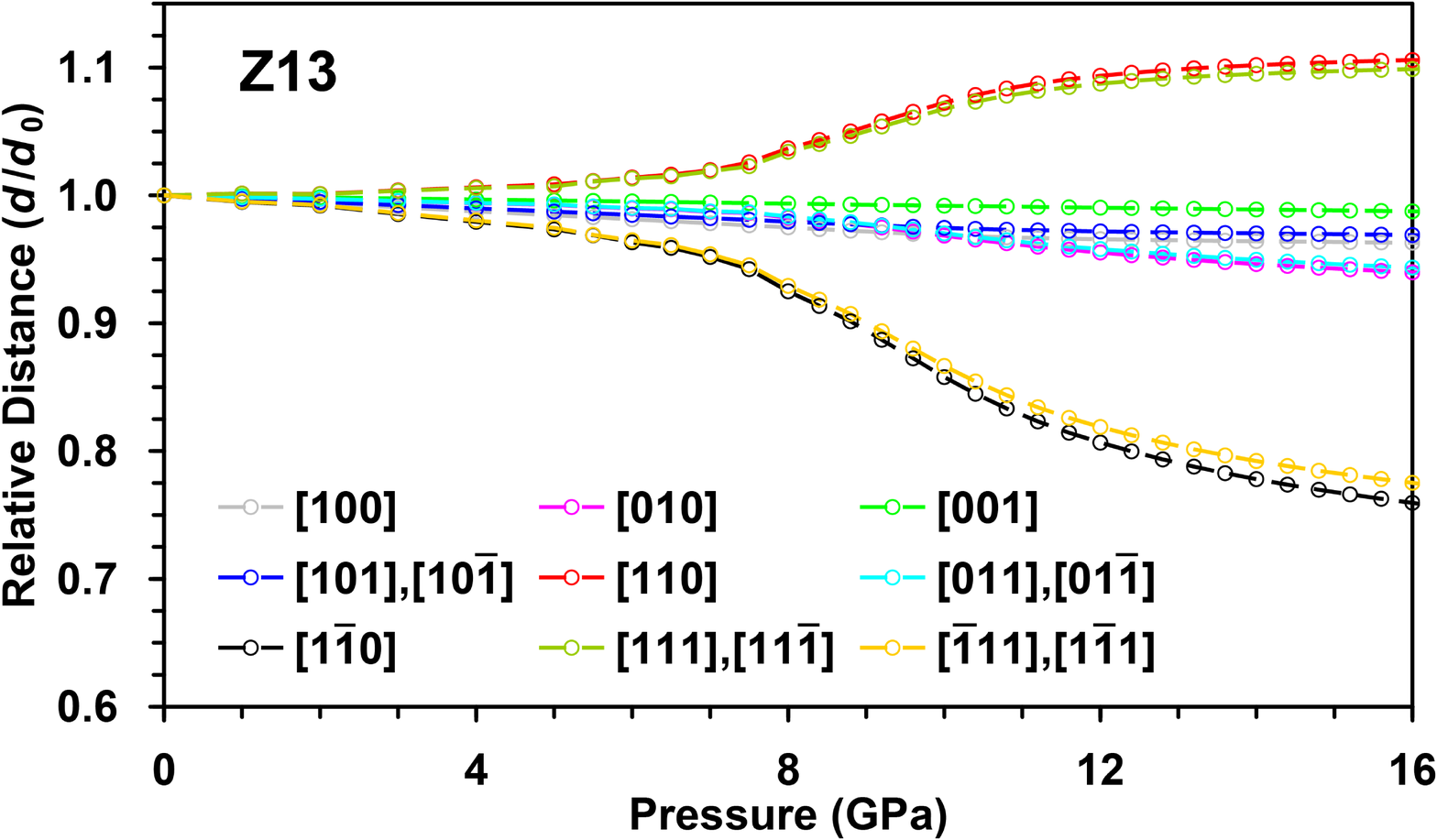}
   \caption{Z13}
   \label{fig:Fig05cd}
\end{subfigure}
\caption{Pressure-dependent lattice distances in thirteen crystal directions of (a) A13 and (b) Z13. In this work, lattice distance indicates the distance between two closest lattice points in the corresponding direction.}
\label{fig:Fig05}
\end{figure}

\begin{figure}[!ht]
\centering
\begin{subfigure}[b]{0.8\textwidth}
   \includegraphics[width=0.8\linewidth]{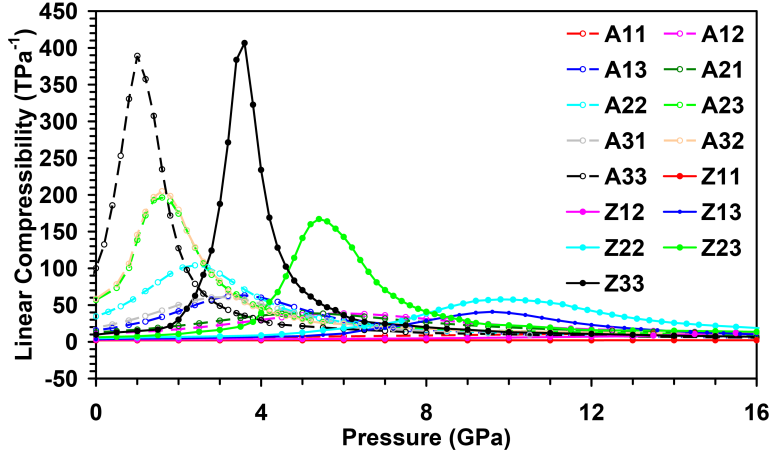}
   \caption{Most positive compressibilities}
   \label{fig:Fig06a} 
\end{subfigure}
\begin{subfigure}[b]{0.8\textwidth}
   \includegraphics[width=0.8\linewidth]{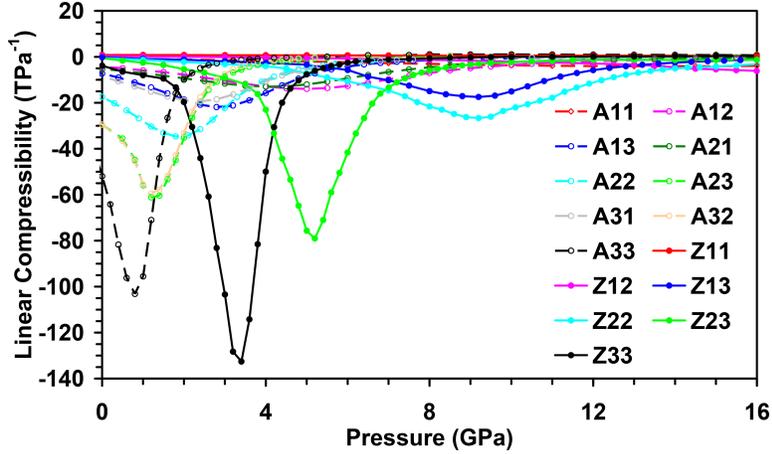}
   \caption{Most negative compressibilities}
   \label{fig:Fig06b}
\end{subfigure}
\caption{Pressure-dependent linear compressibilities of IGN allotropes in the most-positive and most-negative directions. The directions with most positive and negative linear compressibilities are conjugated with each other and in the same surface perpendicular to the pore direction. In the primitive cell, if c ([001]) represents the pore direction, and a ([100]) and b ([010]) indicate two directions parallel to pore sides, like shown in Fig. \ref{fig:Fig05}, then [110] and [1$\bar{1}$0] are the directions with most negative and positive linear compressibilities.}
\label{fig:Fig06}
\end{figure}

Although the unit cell volumes of IGN allotropes decrease with increasing pressure, as required by thermodynamics, their compressibilities are unusually non-monotonic and anisotropic (Figs. \ref{fig:Fig04}-\ref{fig:Fig06}). That is, the compressibilities of all IGNs change dramatically with pressure and are extremely sensitive to the magnitude of applied pressure. For this reason, we do not describe zero-pressure bulk moduli (as typically done for carbon allotropes), but rather discuss the pressure-dependent bulk and linear compressibilities of these phases. As pressure is initially applied from 0 GPa, the bulk compressibilities all increase except Z11 and Z12. For Z11 and Z12, bulk compressibilites decrease slightly in a specific range of pressure (0$-$26 GPa for Z11 and 0$-$6 GPa for Z12) before increasing at high pressures. The bulk compressibility reaches a maximum at a structure-dependent value, and then decreases as pressure is increased further. For Z11, this local maximum compressibility happens near 32 GPa and is insignificant compared with the other structures. Similar to graphite and diamond, the highest bulk compressibility ($\beta{_\text{B,m}}$ in Table \ref{tab:EfCom}) of Z11 occurs at 0 GPa (negative pressures were not considered here), whereas finite pressures for maximum bulk compressibility ($p{_\text{m}}$ in Table \ref{tab:EfCom}) were observed for other IGN allotropes. A general tendency within the same structure type (armchair or zigzag) is that the highest bulk compressibility increases with pore size, while $p_\text{m}$ decreases with increasing pore size (as mentioned above, Z11 is an exception). Differences in the highest bulk compressibilities between different IGN allotropes can vary by orders of magnitude. For example, among all the 15 IGN allotropes, the highest bulk compressibility of A33 is 306 TPa$^{-1}$, however that of Z11 is only 3.6 TPa$^{-1}$.

Jiang et al. \cite{Jiang13} found that only Z22 and Z33 have negative linear compressibilities. Here, we show that this behavior is actually general to the entire IGN family. There are thirteen different linear directions within the primitive cell of a crystal, and for monoclinic structures the principal compression axes are not necessarily coincident with the conventional lattice directions. The linear compressibilities in all directions of IGNs are diverse. With A13 and Z13 as examples (see Fig. \ref{fig:Fig05}), there are three directions ([110],[111], and [11$\bar{1}$]) along which expansion is observed over a certain pressure range. This increase in lattice parameter gives rise to negative linear compressibilitiy (NLC). Meanwhile, the lattice parameters corresponding to other directions decrease with pressure and the resulting linear compressibilities are positive. Each IGN has one direction with a most negative linear compressibility and another with a most positive linear compressibillity (PLC). Similar to bulk compressibility, the linear comepressibilities in the most positive and most negative directions are pressure dependent, and the most positive and negative linear compressibilities are also non-monotonic (Fig. \ref{fig:Fig06}).

\begin{table}[ht]
\caption{Formation enthalpies and extreme properties of carbon allotropes. In this table, $\Delta H$ (eV/atom) denotes formation enthalpy at 0 GPa, $\beta{_\text{B,m}}$ (TPa$^{-1}$) is the highest bulk compressibility and $p_\text{m}$ (GPa) is the corresponding pressure. $\beta{_\text{L,P,m}}$ and $\beta{_\text{L,N,m}}$ denote the maxima of positive and negative linear compressibilities (TPa$^{-1}$), respectively. $p_\text{ir}$ (GPa) (the subscript ir means irreversibly) indicates the pressure range that the IGN irreversibly transforms into a new configuration under cold compression.}
\begin{tabular}{@{}lcccccc@{}}
\hline
Allotrope & $\Delta H$ & $p_\text{m}$ & $\beta{_\text{B,m}}$ & $\beta{_\text{L,P,m}}$ & $\beta{_\text{L,N,m}}$ & $p_\text{ir}$ \\
\hline
Graphite & 0.000 &  0.0 & 299  &      &       & --- \\
Diamond  & 0.139 &  0.0 & 2.3  &      &       & --- \\
A11      & 0.432 & 19.0 & 14.8 & 20.4 & -6.85 & 40-50 \\
A12      & 0.323 &  6.5 & 28.0 & 38.6 & -13.9 & 20-30 \\
A13      & 0.260 &  3.6 & 52.2 & 63.2 & -21.9 & 70-80 \\
A21      & 0.328 &  5.5 & 28.4 & 39.0 & -13.0 & 20-30 \\
A22      & 0.258 &  2.7 & 77.4 & 104  & -34.7 & 70-80 \\
A23      & 0.215 &  1.8 & 161  & 196  & -61.1 & 20-30 \\
A31      & 0.265 &  3.2 & 50.3 & 60.7 & -19.4 & 70-80 \\
A32      & 0.216 &  1.6 & 167  & 205  & -59.5 & 18-20 \\
A33      & 0.185 &  1.2 & 306  & 389  & -103  & 70-80 \\
Z11      & 0.244 &  0.0 & 3.6  & 3.16 & -0.74 & 180-200 \\
Z12      & 0.198 & 19.5 & 11.7 & 18.6 & -8.37 & 30-40 \\
Z13      & 0.163 & 10.0 & 28.5 & 40.8 & -17.5 & 60-70 \\
Z22      & 0.168 & 10.7 & 36.9 & 57.7 & -26.6 & 250-300 \\
Z23      & 0.142 &  5.8 & 114  & 167  & -79.1 & 20-30 \\
Z33      & 0.123 &  3.6 & 283  & 407  & -133  & 200-250 \\
\hline
\end{tabular}
\label{tab:EfCom}
\end{table}

Z33 has the largest PLC and NLC among these 15 IGNs (Table \ref{tab:EfCom}). Their values (PLC: 407 TPa$^{-1}$, NLC: -133 TPa$^{-1}$) pass beyond the reported ``giant'' linear compressibilities in Ag$_3$[Co(CN)$_6$] (PLC: 115  TPa$^{-1}$, NLC: -76 TPa$^{-1}$)\cite{Goodwin08} and Zn[Au(CN)$_2$]$_2$ (PLC: 52 TPa$^{-1}$, NLC: -42 TPa$^{-1}$)\cite{Cairns13}. We noticed that the linear compressibilites of Ag$_3$[Co(CN)$_6$] were obtained in the pressure range of 0-0.19 GPa\cite{Goodwin08} and those of Zn[Au(CN)$_2$]$_2$ were between 0-1.8 GPa\cite{Cairns13}, while the largest linear compressibilities in Z33 were calculated at 3.6 GPa for PLC and 3.4 GPa for NLC. In the pressure range of 0-2.0 GPa, A33 (among 15 IGNs) has the largest linear compressibilites (PLC: 389 TPa$^{-1}$ at 1.0 GPa, NLC: -103 TPa$^{-1}$ at 0.8 GPa). Both the positvie and negative linear compressibilities of Z33 are larger than any previously reported high-pressure ($>$1.0 GPa) values for crystals, despite the fact that none of them exceed the ambient-pressure values (PLC: 430 TPa$^{-1}$, NLC: -260 TPa$^{-1}$) for  CsH$_2$PO$_4$\cite{Prawer85} calculated by Cairns and Goodwin\cite{Cairns15} derived from elastic stiffness components determined by ultrasonic velocity measurements.

\subsection{\label{sec:level2c}Electronic Properties}

Previous reports of the electronic properties of IGN allotropes are in stark contrast. Jiang \textit{et al.}\cite{Jiang13} concluded Z11 is a semiconductor with a band gap between 0.36$-$0.49 eV depending on the type of functional used (Heyd$-$Scuseria$-$Ernzerhof hybrid functionals (HSE06) or Perdew$-$Burke$-$Ernzerhof functionals (PBE)). Chen \textit{et al.}\cite{Chen15}, on the other hand, showed that Z11 is semimetal from both first-principles DFT calculations (PBE) and tight-binding modelling. Here, we confirm that Z11 is indeed a node-line semimetal based on our own DFT-PBE computations, which are in agreement with the results of Chen \textit{et al.}\cite{Chen15}. In addition, we investigated the detailed electronic properties of all 15 IGN allotropes using densities of states (DOS) analysis, one-dimensional electronic band dispersions, and the Fermi surface and band contacts for the semimetallic structures. 

\begin{figure}[!ht]
\centering
\begin{subfigure}[b]{0.8\textwidth}
   \includegraphics[width=0.8\linewidth]{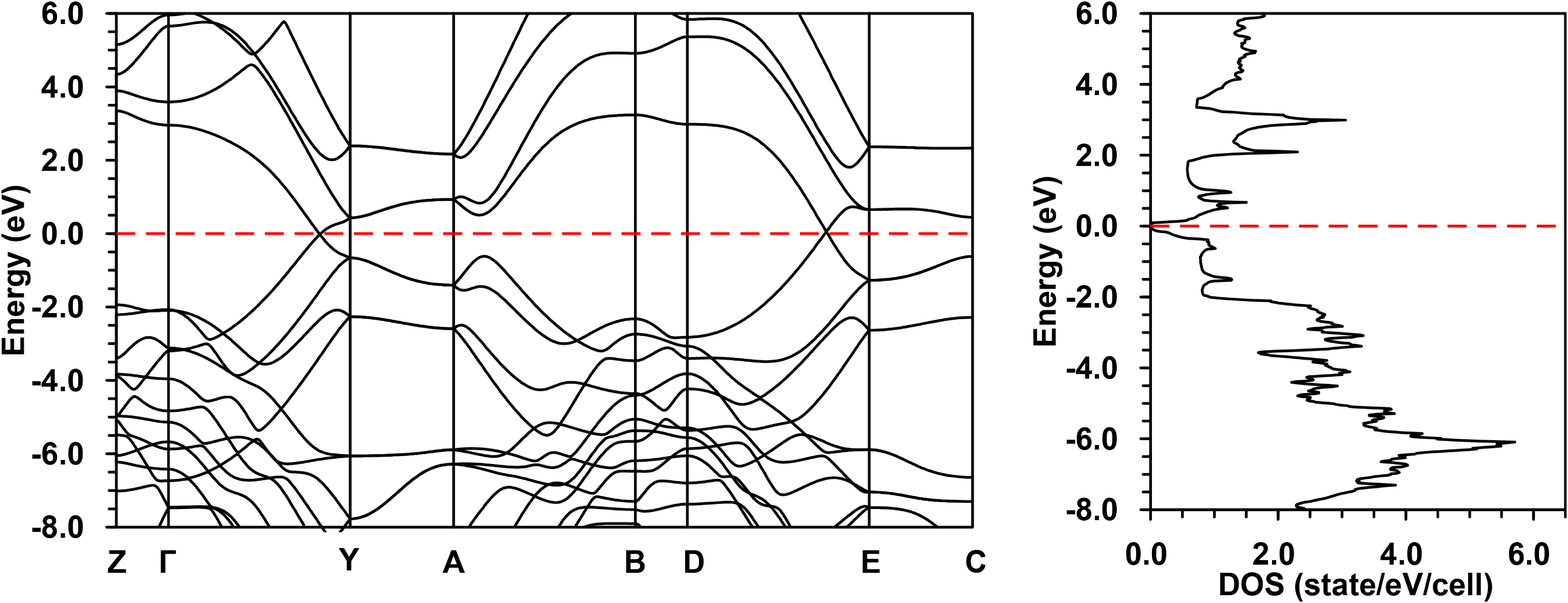}
   \caption{Z13}
   \label{fig:Fig07a} 
\end{subfigure}
\begin{subfigure}[b]{0.8\textwidth}
   \includegraphics[width=0.8\linewidth]{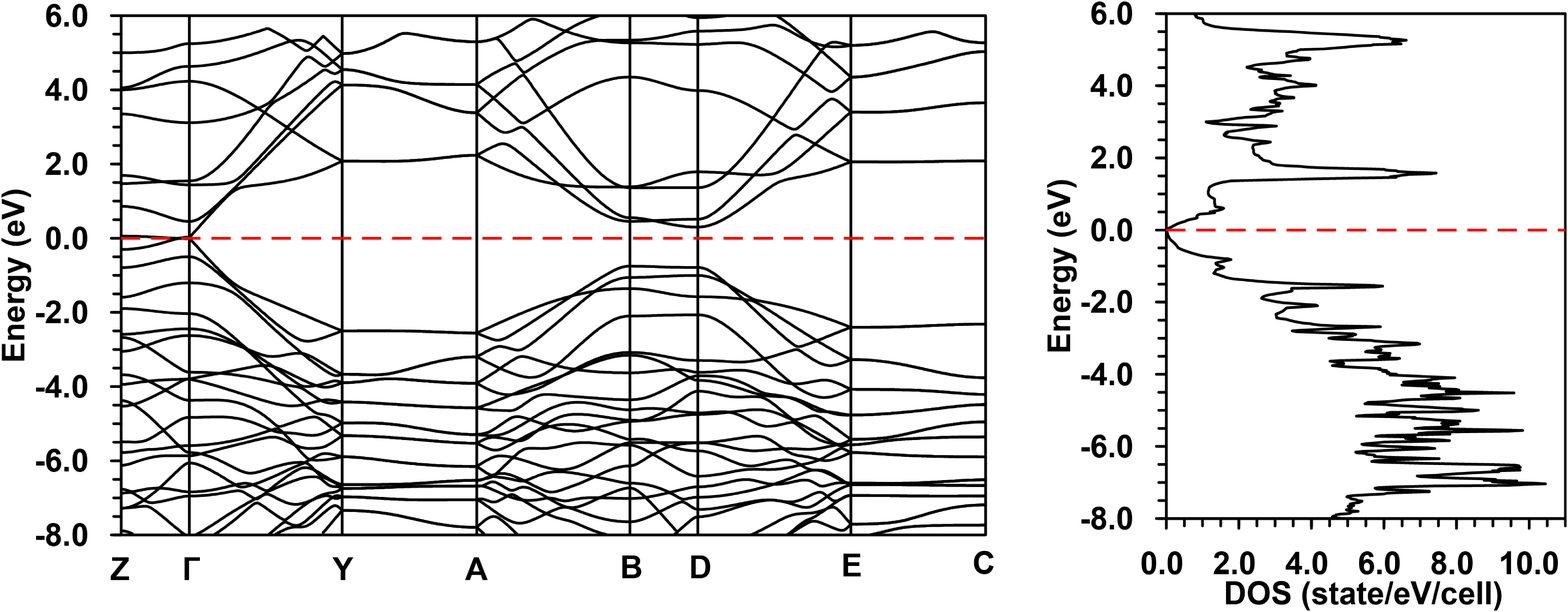}
   \caption{A13}
   \label{fig:Fig07b}
\end{subfigure}
\begin{subfigure}[b]{0.8\textwidth}
   \includegraphics[width=0.8\linewidth]{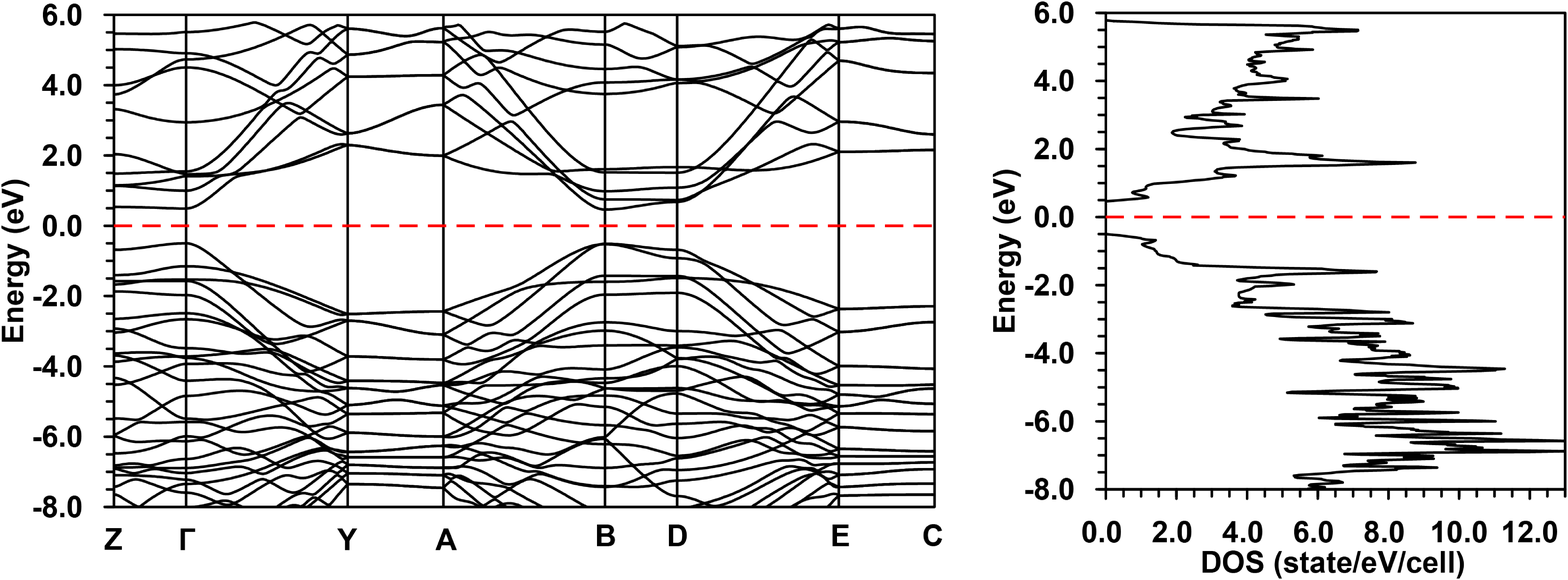}
   \caption{A33}
   \label{fig:Fig07c}
\end{subfigure}
\caption{Electronic band dispersion relations and densities of states for (a) Z13, (b) A13 and (c) A33.}
\label{fig:Fig07}
\end{figure}

At the Fermi energy level, we found that the density of states in Z13, for example, is very small (on the order of 10$^{-3}$ states/cell/eV) but not zero, and the highest valence band and the lowest conduction band contact in the Gamma to Y and D to E directions (Fig. \ref{fig:Fig07}a). Since the contact points are not located at high-symmetry points, they could be easily missed without using dense k-point grids. Similar situations exist for all of the Z-type and some A-type (A11,A12,A13,A21,A31) IGNs (Fig. \ref{fig:Fig07}b and Figs. S12-S15 in the supporting information). Thus, all of the Z-type and five of the A-type IGNs are semimetals (no band gap, but vanishingly small density of states at the Fermi level). In contrast to this behavior, we found that some of the large-pore, A-type IGNs are semiconducting. For A33, we found a band gap of 0.48 eV in the band dispersion relations, which was also confirmed using the density of states. This is similar to the findings of Jiang et al.\cite{Jiang13}, but the magnitude of the gap is different. We attribute the difference (including the finding that Z11 is actually a semimetal) to a finer sampling of the Brillioun zone. Similar to A33, A22, A23 and A32 are also found to be semiconductors with band gaps of 0.92, 0.96, and 0.66 eV, respectively, at the DFT (PBE) level (Figs. S10 and S11). 

\begin{figure}[!ht]
\centering
\includegraphics[width=0.8\linewidth]{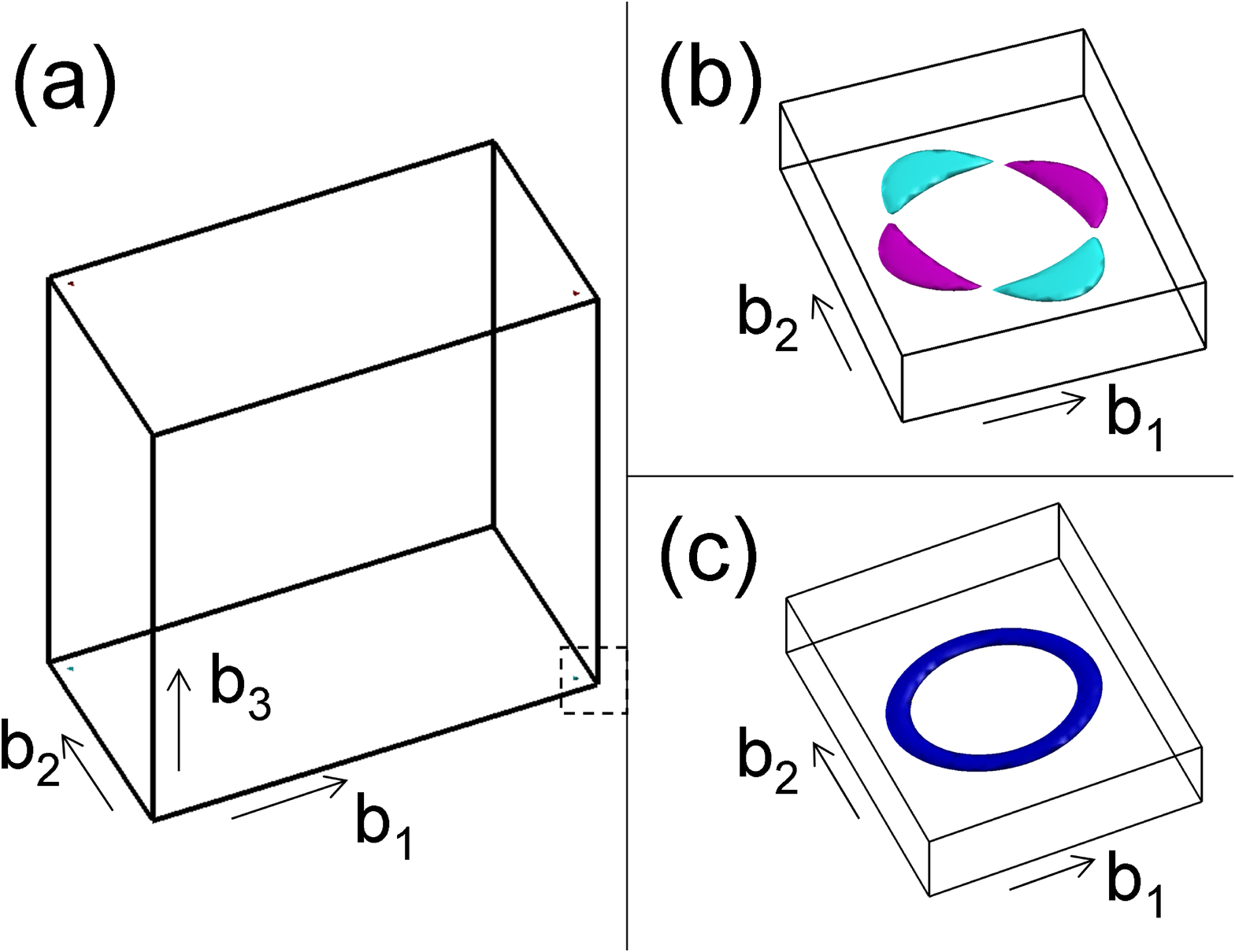}
\caption{Fermi surface (a,b) and isoenergy difference surface (c) in A13. (b) is a gamma-point-centered local representation of (a) (indicated by the dashed rectangle), with a k-point density 2500 times larger than in (a). No Fermi surface can be found in (a) other than the space shown in (b). (c) represents an isoenergy difference surface (0.01 eV) between the highest-energy valance band and the lowest-energy conduction band. b$_1$, b$_2$ and b$_3$ in this and following figures indicate the reciprocal lattice directions corresponding to a, b and c of Bravais lattices}
\label{fig:Fig08} 
\end{figure}

The semimetallic structures all show band dispersion features similar to graphene. We extracted the Fermi surfaces for all semimetallic allotropes to further analyze their electronic structures. This process requires an extremely dense k-point grid in reciprocal space in order to obtain a clear picture of the Fermi surface. With A13 as an example of the A-type IGNs, we found that the Fermi surface exists within non-connected local areas (Fig. \ref{fig:Fig08}a). Very fine k-point grids are required to delineate the Fermiology. We used a k-point mesh of 28$\times$16$\times$48, corresponding to a spacing of 0.004, 0.004 and 0.0025 $\text{Bohr}^{-1}$ in the $b_1$, $b_2$ and $b_3$ directions, respectively. The Fermi surface became clearer (Fig. \ref{fig:Fig08}b) when we used a spacing of 0.0004, 0.0004 and 0.0001 $\text{Bohr}^{-1}$. The Fermi surface is comprised of four thin Fermi arcs, similar to the Fermi arcs observed in the Weyl semimetal TaAs \cite{Lv15} (Fig. S16 in the supporting information). The isoenergy surface, derived from the energy difference between the highest-energy valance band and the lowest-energy conduction band, looks like a circular loop in reciprocal space, indicating contact points (nodes where the energy difference between bands is zero). Within this contact loop, there are 4 points with band energies that are exactly the same as Fermi energy. Thus, A33 can be described as a node-loop semimetal.

\begin{figure}[!ht]
\centering
\includegraphics[width=0.8\linewidth]{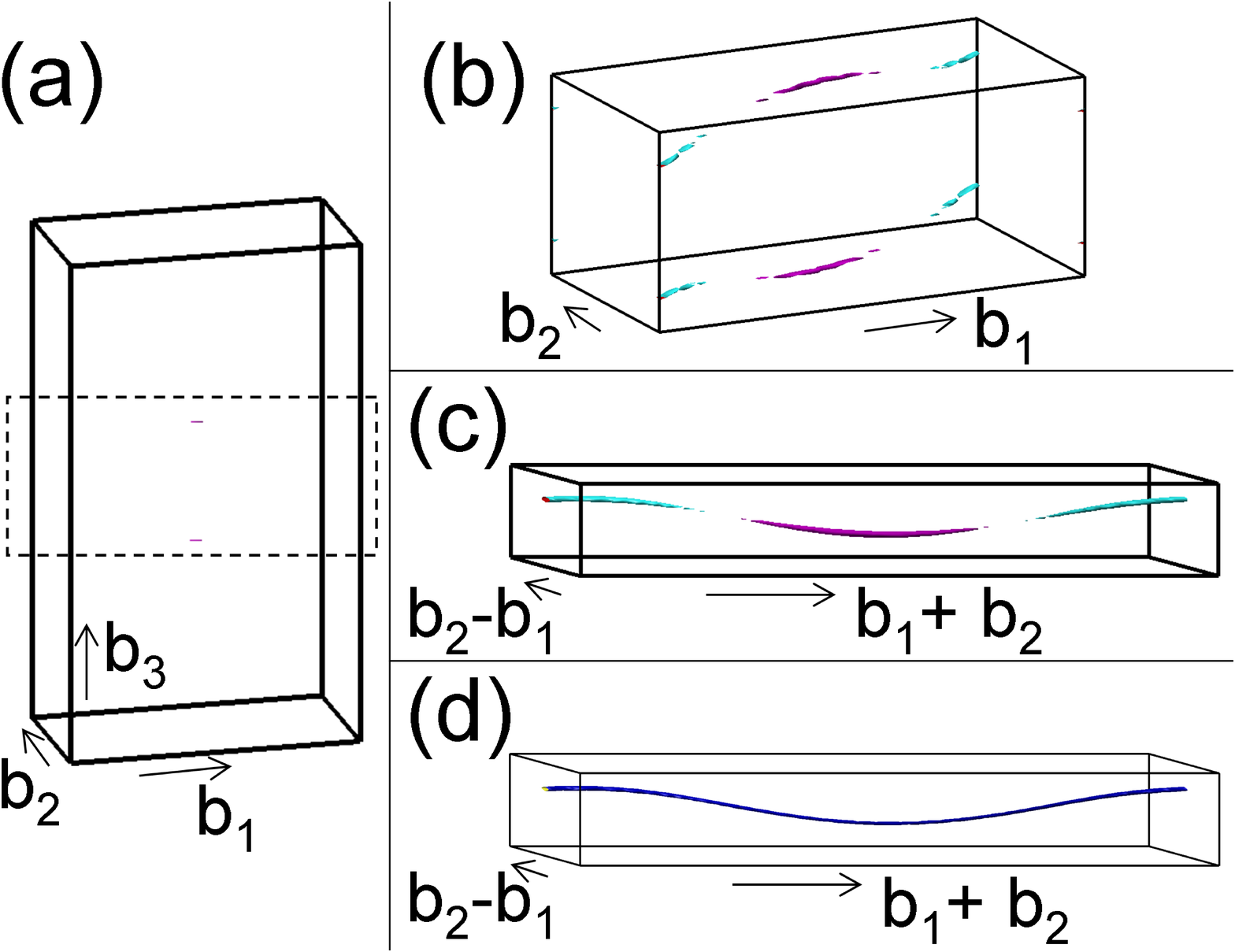}
\caption{Fermi surface (a,b,c) and isoenergy difference surface (d) for Z13. (b) is the local space enclosed by the dashed rectangle in (a) that includes the Fermi surface, (c) represents the lower portion of (b), but from different lattice directions.  The k-point density in (c) is 8 times larger than in (b) and 250 times larger than in (a). (d) is the isoenergy difference surface (0.05 eV) between the highest-energy valance band and the lowest-energy conduction band.}
\label{fig:Fig09} 
\end{figure}

Looking at Z13 as an example for Z-type IGNs, the Fermi surface was also very unclear using a spacing of 0.005 $\text{Bohr}^{-1}$ in all $b_1$, $b_2$ and $b_3$ directions (Fig. \ref{fig:Fig09}a). It became clearer using a spacing of 0.002, 0.002 and 0.001 $\text{Bohr}^{-1}$, but still displayed an intermittent pattern (Fig. \ref{fig:Fig09}b). Using an even smaller spacing of 0.001, 0.001 and 0.0005 $\text{Bohr}^{-1}$ (Fig. \ref{fig:Fig09}c and d), we conclude that the Fermi surface of Z13 is actually connected. The Fermi surface of Z13 is formed by two symmetric lines. Each line is connected by Fermi arcs, also similar to the Fermi arcs in the Weyl semimetals TaAs \cite{Lv15} (Fig. S16 in the supporting information). Different from the case of A13, the isoenergy difference surface of Z13 looks like hollow lines, which indicates that the contact points form two lines in reciprocal space. Within these contact lines, there are four points whose band energies are exactly the same as the Fermi energy. Thus, Z13 can be described as a node-line semimetal.

\begin{figure}[!ht]
\centering
\includegraphics[width=0.8\linewidth]{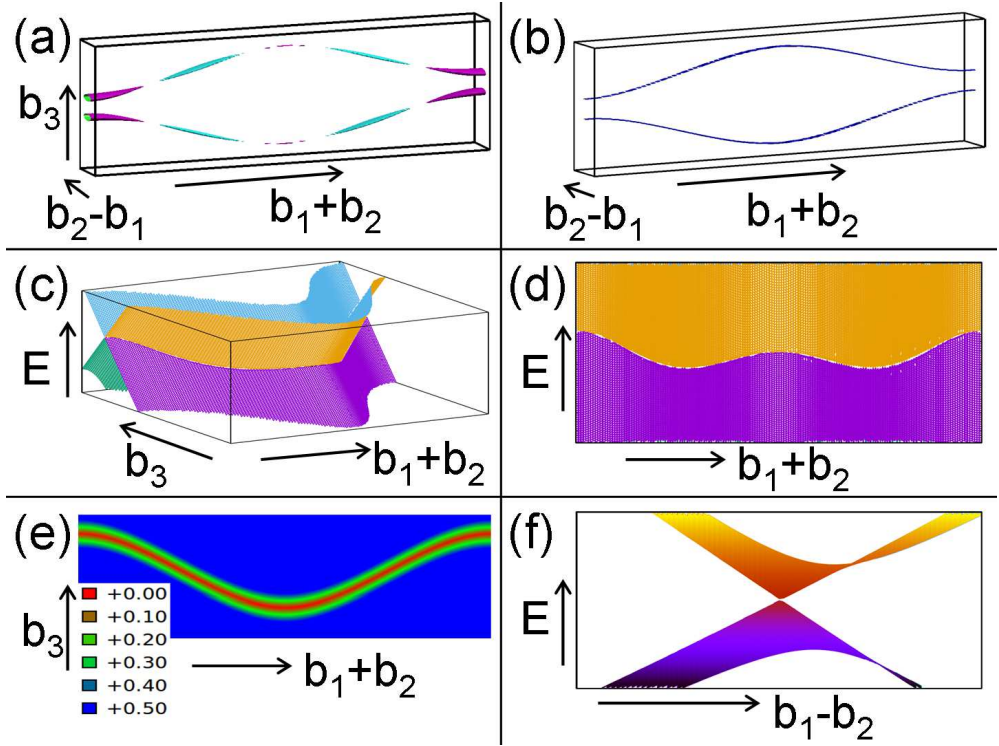}
\caption{Fermi surface (a), isoenergy difference surface (b), and two-dimensional electronic band dispersion in Z13. (a) is the Fermi surface using k-point grid intervals of 0.001, 0.001, and 0.0005 $\text{Bohr}^{-1}$ in the $b_1+b_2$, $b_2-b_1$ and $b_3$ directions, respectively. (b) is an isoenergy difference surface (0.05 eV) between the highest-energy valance band and the lowest-energy conduction band. (c) and (d) represent two-dimensional electronic band dispersion viewed from different projections in a plane using a fixed value in the $b_2-b_1$ direction. (e) is the band-energy difference within the same plane of reciprocal space used in (c) and (d). (f) represents the two-dimensional electronic band dispersion in a plane with fixed value in the $b_1+b_2$ direction.}
\label{fig:Fig10} 
\end{figure}

Using the same procedure described above for A13 and Z13, we found that all Z-type IGNs, as well as A21, are node-line semimetals, while A11, A12, A13, and A31 are node-loop semimetals (Table \ref{tab:MetEg}, Fig. \ref{fig:Fig10}, and Figs. S17-S24 in the supporting information).

Now we come back to Z11, the first IGN allotrope suggested to be  semimetallic\cite{Chen15}. Similar to Z13, the Fermi surface for Z11 is also formed by two symmetric lines, and the band contact points also form two lines in reciprocal space (Fig. \ref{fig:Fig10}a and b). In addition to the isoenergy difference surface (an indirect way of showing band contact properties), we directly show that the bands contact on a line by examining the two-dimensional energy band dispersion (Fig. \ref{fig:Fig10}c and d). Although we can also see that the bands contact in the one-dimensional dispersion plot (Fig. S14 in the supporting information), we can only observe isolated single points. In general, the whole contact line cannot be visualized by way of a two-dimensional dispersion plot, but by taking into account the crystallographic symmetry, the whole contact line in Z11 can be observed in a plane with fixed values in the $b_2-b_1$ direction (Fig. \ref{fig:Fig10}c-f). The electronic bands of Z11 are linearly dispersive (typical characteristic for Dirac and Weyl semimetals around the Dirac or Weyl points) in planes with fixed values in the $b_1+b_2$ direction (Fig. \ref{fig:Fig10}f). Thus, we also demonstrate that Z11 is a node-line semimetal in both indirect and direct ways.

\begin{table}[ht]
\caption{Electronic properties of IGN allotropes. In this table, nodes (for semimetals) indicate the shape of contact points between the lowest-energy conduction band and the highest-energy valence band, while $E_g$ (eV) is the band gap for semiconductors.}
\begin{tabular}{@{}lccc@{}}
\hline
Allotrope & metallicity & nodes & $E_g$ \\
\hline
A11      & semimetal     & loop  & ---  \\
A12      & semimetal     & loop  & ---  \\
A13      & semimetal     & loop  & ---  \\
A21      & semimetal     & lines & ---  \\
A22      & semiconductor & ---   & 0.92 \\
A23      & semiconductor & ---   & 0.96 \\
A31      & semimetal     & loop  & ---  \\
A32      & semiconductor & ---   & 0.66 \\
A33      & semiconductor & ---   & 0.48 \\
Z11      & semimetal     & lines & ---  \\
Z12      & semimetal     & lines & ---  \\
Z13      & semimetal     & lines & ---  \\
Z22      & semimetal     & lines & ---  \\
Z23      & semimetal     & lines & ---  \\
Z33      & semimetal     & lines & ---  \\
\hline
\end{tabular}
\label{tab:MetEg}
\end{table}

\section{\label{sec:level1d}Conclusion}

In this work, we demonstrate that interpenetrating graphene networks are metastable pure carbon allotropes with relatively low formation enthalpies ($0.1-0.5$ eV/atom). Among all 15 IGN allotropes with mechanical stability at 1 atm, Z33 is the most energetically favorable IGN allotrope at $P<$1.7 GPa and Z11 is the most energetically favorable one at pressures $P>$9.7 GPa. Between 1.7$<P<$9.7, Z13 is the most energetically favorable.

Non-monotonic bulk and negative linear compressibilities are two typical characteristics of IGNs, which are unusual compared with crystals of other carbon allotropes and most materials in general. The highest bulk compressibilities and the largest negative linear compressibilities depend on the specific structures.

All Z-type IGNs are node-line semimetals. For A-type IGNs, A22, A23, A32 and A33 are semiconductors with band gaps of 0.92, 0.96, 0.66, and 0.48 eV, respectively. A21 is a node-line semimetal, while A11, A12, A13, A31 are all node-loop semimetals.

These novel carbon allotropes offer attractive multifunctional properties that might see experimental realization through synthetic strategies such as metal removal from high-pressure MC$_6$ carbides.

\section{Acknowledgments}
\label{acknowledgments}
This work is supported by DARPA under grant No. W31P4Q1310005. Our DFT computations were performed on the supercomputer Copper of DoD HPCMP Open Research Systems under project No. ACOMM35963RC1 as well as on XSEDE supercomputers and local clusters (mw and memex) of Carnegie Institution for Science. REC is supported by the Carnegie Institution for Science and by the European Research Council Advanced Grant ToMCaT. The authors wish to thank D. Vanderbilt, J. Liu and I. Naumov for their fruitful discussions and helpful comments.

\clearpage
\bibliographystyle{apsrev4-1}

\end{document}